\documentclass[journal]{IEEEtran}

\usepackage{amsmath,amsfonts}
\usepackage{algorithmic}
\usepackage{array}
\usepackage[caption=false,font=normalsize,labelfont=sf,textfont=sf]{subfig}
\usepackage{textcomp}
\usepackage{stfloats}
\usepackage{url}
\usepackage{verbatim}
\usepackage{graphicx}
\usepackage[utf8]{inputenc}
\usepackage[numbers,sort&compress]{natbib}
\usepackage{booktabs,array,makecell}
\usepackage{stfloats}   
\usepackage{float}
\usepackage{rotating}
\usepackage{csquotes}
\usepackage{amsmath}
\usepackage{bm}
\usepackage{tabularx}
\usepackage{stfloats}
\usepackage{threeparttable}
\usepackage{url}
\usepackage{soul}
\usepackage{threeparttable}
\soulregister\cite7
\soulregister\ref7
\soulregister\pageref7
\usepackage{amssymb}
\usepackage{soul}
\usepackage{footnote}
\usepackage{colortbl}
\usepackage{soul}
\usepackage{multirow}
\usepackage{pifont}
\usepackage{algorithmic}
\usepackage{booktabs,xcolor,colortbl}
\usepackage{color}
\usepackage{alltt}
\usepackage{hyperref}
\usepackage{enumerate}
\usepackage{siunitx}
\usepackage{epstopdf}
\usepackage{bbding}
\usepackage{pbox}
 \usepackage{amsmath}  
 \usepackage{amssymb}   
 \usepackage{amsthm}
 \usepackage[dvipsnames]{xcolor}
 
\usepackage{algorithm}
\usepackage{xcolor}

\newcommand{\probP}{\text{I\kern-0.15em P}}
\newcommand{\probE}{\text{I\kern-0.15em E}}

\def\BibTeX{{\rm B\kern-.05em{\sc i\kern-.025em b}\kern-.08em
    T\kern-.1667em\lower.7ex\hbox{E}\kern-.125emX}}
\usepackage{balance}
\begin{document}
\title{Directionally Supervised Neuromorphic Protection for Distribution Systems Using Local Measurements}
\author{Saurabh Prabhakar, \textit{Member, IEEE},  Yubo Song,  \textit{Senior Member, IEEE}, Pascal Schirmer, \textit{Member, IEEE}, Bijaya Ketan Panigrahi, \textit{Fellow, IEEE}, and Subham Sahoo, \textit{Senior Member, IEEE}
}

\maketitle

\begin{abstract}
Reliable protection under varying power system operating conditions requires rapid fault detection while preserving selectivity against external faults and non-fault disturbances. This requirement is particularly challenging in communication-free schemes, where relays must determine trip conditions solely from local electrical measurements. To address this challenge, this paper proposes a directionally supervised neuromorphic protection scheme that integrates disturbance-responsive leaky integrate-and-fire dynamics with a direction-dependent firing threshold to distinguish internal faults from reverse-direction events. The unified relay architecture combines local disturbance detection, directional discrimination, and spike-based trip generation without requiring inter-relay communication. Its performance is evaluated under grid-connected and islanded conditions using a ring microgrid and the IEEE 13-node distribution feeder. Across 240 evaluated fault scenarios, the framework achieves median operating times of 5.4--7.6~ms across the two networks, with 78\% of cases operating within 10~ms and 95\% within 20~ms. It also maintains protection security during load disturbances, while fault-resistance analysis confirms its response under high-resistance faults. Controller-hardware-in-the-loop (CHIL) experiments using an RTDS simulator and an external dSPACE controller demonstrate trip-command assertion in approximately 10~ms and 5~ms for the tested single-line-to-ground and three-phase-to-ground faults, respectively. The results establish real-time feasibility within the evaluated conditions and characterize operating-time variability across fault types and resistance levels for different network configurations.
\end{abstract}


\section{Introduction}
\IEEEPARstart{T}{he} growing penetration of inverter-based distributed energy resources (DERs) is fundamentally altering fault characteristics in distribution networks and microgrids \cite{narang2021overview}. Limited and controller-dependent fault currents, bidirectional power flows, and transitions between grid-connected and islanded operation pose significant challenges to conventional protection schemes that rely on predictable fault-current magnitudes and directions \cite{hooshyar2017microgrid, manson2021practical, beheshtaein2019dc}. These challenges are particularly pronounced in inverter-dominated microgrids, where fault responses are strongly influenced by converter control strategies and prevailing operating conditions \cite{aliyu2025comprehensive}. Consequently, achieving rapid fault detection while preserving selectivity between internal and external faults remains a fundamental requirement for reliable protection~\cite{barranco2023adaptive}.

Directional protection is essential in these networks because it enables relays to discriminate faults according to their direction relative to the protected zone. Conventional directional elements typically rely on voltage--current relationships, sequence components, or impedance-based characteristics to determine fault direction. However, reduced fault-current magnitudes, modified voltage--current phase relationships, and voltage depression can impair directional assessment in inverter-based microgrids \cite{hooshyar2017new, jia2019influence}. Although advanced techniques, such as harmonic-based directional relaying, have been developed to improve protection performance under these conditions, their effectiveness depends on the availability and characteristics of the electrical signatures used for directional discrimination \cite{el2020harmonic, li2014traveling, paladhi2024reliable}. These limitations highlight the need for protection mechanisms that respond rapidly to local electrical disturbances while maintaining dependable directional selectivity.

Communication-assisted protection enhances selectivity by exchanging measurements or protection information between network terminals \cite{zhou2024sampled}. However, its operation remains dependent on the availability and reliability of the underlying communication infrastructure. Communication-free, or non-pilot, protection eliminates this dependency but requires each relay to discriminate faults using only the measurements available at its own terminal \cite{mohammadhassani2023nonpilot, armaghani2026novel, beikbabaei2024machine}. Accordingly, a relay must distinguish a forward internal fault from a reverse-direction event without access to remote measurements, while maintaining security against external faults and non-fault disturbances, including load changes. \textit{The fundamental challenge is therefore to achieve rapid and directionally selective protection using exclusively local electrical measurements under the variable operating conditions of inverter-dominated networks.}

Neuromorphic computing offers an alternative paradigm for processing electrical disturbances through event-driven neuronal dynamics. In particular, spiking neural networks (SNNs) encode input variations as discrete spikes, while leaky integrate-and-fire (LIF) neurons integrate incoming excitation and generate spikes when their membrane potential reaches a prescribed firing condition\cite{roy2019towards, tavanaei2019deep}. These characteristics provide a potential foundation for rapid protection decisions without exclusive reliance on conventional phasor estimation. Recent studies have investigated neuromorphic principles for microgrid protection, demonstrating the potential of spike-based mechanisms for disturbance detection and fault localization without conventional inter-relay communication \cite{diao2024neuromorphic, kamoona2026robust}. Complementary research has highlighted the effectiveness of localized, event-driven mechanisms in enhancing the responsiveness and resilience of microgrid operation, although their focus is on cyberattack-resilient control rather than directional fault protection \cite{sahoo2020event}.

Nevertheless, event-driven neuronal dynamics alone cannot ensure directional selectivity, particularly in multi-source networks characterized by bidirectional power flow. The motivation for neuromorphic protection is also rooted in the electrical dynamics of inverter-interfaced distributed energy resources (DERs), whose energy-storage and resistive characteristics can be represented through an equivalent leaky integrate-and-fire (LIF) model, as demonstrated in our previous work~\cite{prabhakar2026synapse}. Building on this physical correspondence, this paper proposes a directionally selective neuromorphic protection framework that integrates local, disturbance-responsive LIF dynamics with a direction-dependent firing threshold. The proposed framework is evaluated under grid-connected and islanded operating conditions using a multi-relay ring microgrid and the IEEE 13-node feeder, with emphasis on operating time, directional selectivity, and security against load changes.

The primary contribution of this work is the integration of local sequence-domain directional discrimination into an event-driven neuromorphic protection framework, enabling selective fault detection using local measurements without relying on remote electrical information. Building on the communication-free neuromorphic protection principle established in our previous work~\cite{prabhakar2026synapse}, which focused on islanded operation, this study introduces directional supervision and extends the evaluation to both grid-connected and islanded conditions.

The specific contributions of this paper are as follows:
\begin{enumerate}
    \item A local directional-supervision mechanism that combines incremental sequence-domain information with disturbance-responsive neuronal dynamics to distinguish forward and reverse disturbances.

    \item A relay-level protection formulation integrating disturbance activation, directional permission, adaptive membrane integration, and trip generation within a unified decision architecture.

    \item Real-time evaluation across two network configurations, including fault-resistance-dependent operating times and external-controller tests of fault response and load-change security.
\end{enumerate}

The remainder of this paper is organized as follows. Section~II reviews the neuromorphic computing principles for power system protection, and Section~III details the proposed directionally selective framework and its formulation. Section~IV evaluates the protection response under representative fault and non-fault scenarios, while Section~V presents the controller-hardware-in-the-loop validation and the quantitative performance across fault types, networks, and fault resistances. Section~VI concludes the paper.

\section{Neuromorphic Computing Foundations for System Protection}

Neuromorphic computing emulates biological neuronal processing, in which input stimuli are represented by discrete spikes and integrated over time to generate an output response. The conversion of continuous-valued inputs into spikes is commonly achieved through rate coding or latency coding\cite{tavanaei2019deep}. In rate coding, the input magnitude is represented by the frequency of spikes generated within a specified time window. For a normalized input $x\in[0,1]$, the firing rate can be expressed as
\begin{equation}
r(x)=r_{\min}+(r_{\max}-r_{\min})x,
\label{eq:rate}
\end{equation}
where $r_{\min}$ and $r_{\max}$ denote the minimum and maximum encoding rates, respectively. The corresponding number of spikes within an observation window $T_w$ is approximately
\begin{equation}
N_x(T_w)\approx r(x)T_w.
\label{eq:rate_spikes}
\end{equation}
Thus, larger input magnitudes produce higher spike frequencies, whereas smaller inputs produce fewer spikes.

In contrast, latency coding represents the input magnitude through the time interval between stimulus presentation and the first generated spike. A commonly used inverse relationship is
\begin{equation}
\tau(x)=
\operatorname{clip}
\left(
\frac{K_{\tau}}{x},
\tau_{\min},
\tau_{\max}
\right),
\qquad x>0,
\label{eq:latency}
\end{equation}
where $K_{\tau}$ is the latency-scaling coefficient, and $\tau_{\min}$ and $\tau_{\max}$ define the encoding limits~\cite{mostafa2017supervised}. Consequently, stronger inputs produce shorter spike latencies, while weaker inputs produce longer latencies. Unlike rate coding, latency coding can represent input magnitude through the timing of a single spike, reducing the need for repeated spike generation within an observation window. These encoding principles provide different mechanisms for representing input magnitude and temporal information~\cite{sahoo2025spike}.

The LIF model provides a computationally efficient representation of neuronal integration, membrane leakage, and threshold-dependent firing. Its dynamics are expressed as
\begin{equation}
\tau_m\frac{dV_m(t)}{dt}
=
-\left[V_m(t)-V_{\mathrm{rest}}\right]
+
R_m I_{\mathrm{in}}(t),
\label{eq:lif_dynamics}
\end{equation}
where $V_m(t)$ is the membrane potential, $V_{\mathrm{rest}}$ is the resting potential, $\tau_m$ is the membrane time constant, $R_m$ is the membrane resistance, and $I_{\mathrm{in}}(t)$ represents the input current generated by the encoded spikes. The LIF dynamics also admit an equivalent RC-circuit interpretation, providing a physical correspondence with the energy-storage and resistive characteristics of inverter-interfaced DERs~\cite{prabhakar2026synapse}. The membrane potential accumulates incoming excitation while decaying towards its resting state, allowing persistent or closely spaced inputs to produce greater accumulation than isolated inputs.

A neuronal spike is generated when the membrane potential reaches the firing threshold $\Theta$. The output is defined as
\begin{equation}
s(t)=
\begin{cases}
1, & V_m(t)\geq\Theta,\\
0, & V_m(t)<\Theta.
\end{cases}
\label{eq:spike_generation}
\end{equation}
Following firing, the membrane potential is updated through a reset mechanism. In a subtractive-reset formulation,
\begin{equation}
V_m(t^+)=V_m(t^-)-\Theta,
\label{eq:lif_subtractive_reset}
\end{equation}
where $V_m(t^-)$ and $V_m(t^+)$ denote the membrane potentials immediately before and after firing, respectively. This mechanism removes the threshold-crossing component while retaining residual membrane excitation for subsequent integration~\cite{rueckauer2017conversion}.

\begin{figure*}[t]
  \centering
  \includegraphics[width=0.9\textwidth]{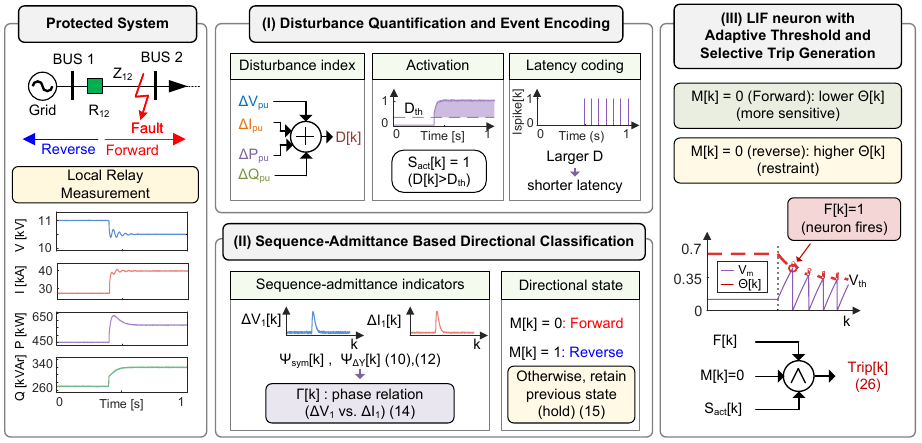}
  \caption{Proposed directionally selective neuromorphic protection framework: Rated-base normalization forms the disturbance index $D[k]$, which drives latency encoding ($\tau_{\mathrm{lat}}\propto 1/D[k]$); a sequence-admittance index $\Gamma[k]$ sets the directional state $M[k]$; and a LIF neuron with a collapsible adaptive threshold $\Theta[k]$ trips only for internal/forward faults.}
  \label{fig1}
\end{figure*}

\begin{figure*}[t]
  \centering
  \includegraphics[width=0.9\textwidth]{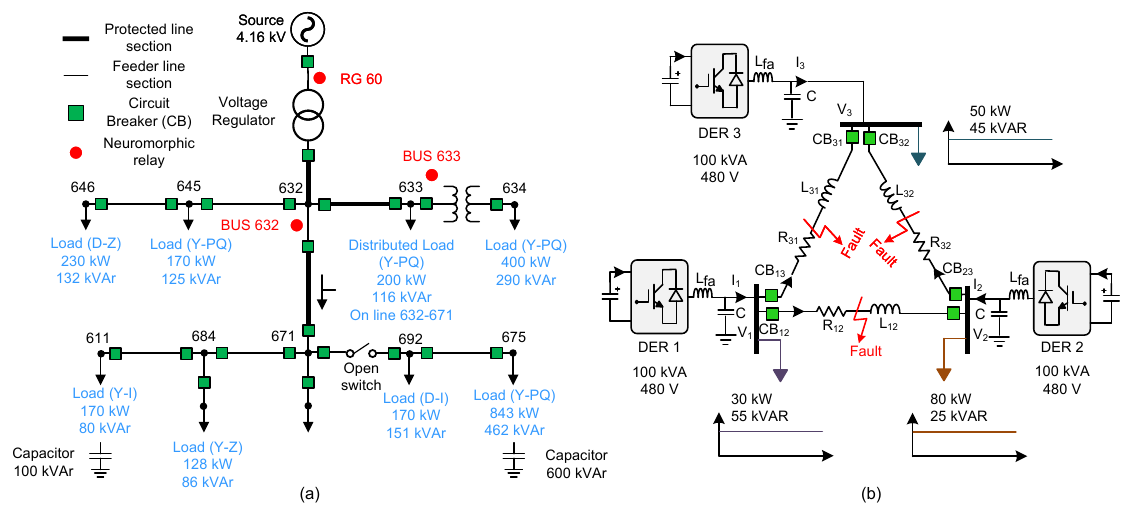}
  \vspace{-6pt}
  \caption{Test networks used for protection validation: (a) the benchmark IEEE 13-node radial feeder, and (b) a three-DER islanded ring microgrid.}
  \label{fig2}
\end{figure*}

\section{Proposed Directionally Selective Neuromorphic Protection Framework}

The proposed framework comprises three interconnected stages, as illustrated in Fig.~\ref{fig1}. First, local voltage, current, active-power, and reactive-power measurements are normalized using their rated bases to construct a multi-domain disturbance index, $D[k]$, which governs event activation and latency-based spike encoding. Second, positive-sequence voltage and current increments are processed through sequence-admittance indicators to determine the signed directional index, $\Gamma[k]$, and the corresponding directional state, $M[k]$. Third, the encoded disturbance spikes are integrated by an LIF neuron, whose adaptive firing threshold, $\Theta[k]$, accounts for disturbance severity and direction. The final trip decision requires simultaneous satisfaction of neuronal firing, disturbance activation, and forward directional permission. The mathematical formulation of these stages and their integration into the protection mechanism are presented in the following subsections.

\subsection{Objective and Selectivity Preliminaries}

Each relay is assigned a forward current reference and a primary protected line section, as illustrated in Fig.~\ref{fig1}. Directional discrimination determines whether the measured disturbance is consistent with the relay's forward or reverse reference direction. However, a forward directional decision alone does not establish whether the fault lies within the primary protected section; line-section selectivity must also be assessed based on network topology and relay assignments.

For each fault scenario, the expected primary relay set is determined from the network topology and breaker locations. Dependability is evaluated based on the operation of the required primary relays, whereas security is assessed through the restraint of relays not assigned to clear the fault. Any intentional coordinated backup operation is considered separately.

\subsection{Local Measurement Normalization and Disturbance Index}

To enable consistent disturbance assessment across networks with different nominal ratings, locally measured voltage, current, active power, and reactive power are expressed as normalized deviations:
\begin{equation}
\Delta X_{\mathrm{pu}}[k]
=
\frac{|x[k]-x_{\mathrm{ss}}|}{x_{\mathrm{ss}}},
\quad
x\in\{V,I,P,Q\},
\label{eq:normalized_measurements}
\end{equation}
\noindent where $x[k]$ is the measured magnitude at sampling instant $k$, and $x_{\mathrm{ss}}$ is its rated steady-state reference. Rated phase quantities are used for voltage and current, while the rated apparent power $S_{\mathrm{rated}}$ serves as the common reference for active and reactive power. Thus, all four deviations are dimensionless.

The normalized deviations are aggregated using equal weighting to obtain the scalar disturbance index:
\begin{equation}
D[k]
=
\Delta V_{\mathrm{pu}}[k]
+
\Delta I_{\mathrm{pu}}[k]
+
\Delta P_{\mathrm{pu}}[k]
+
\Delta Q_{\mathrm{pu}}[k].
\label{eq:disturbance_index}
\end{equation}

This magnitude-based aggregation avoids empirically fitted weighting coefficients and retains the contribution of all four measurement channels irrespective of deviation sign. Event-driven encoding is activated when the index exceeds the common threshold $D_{\mathrm{th}}$:
\begin{equation}
\mathcal{E}[k]
=
\begin{cases}
1, & D[k]\geq D_{\mathrm{th}},\\
0, & D[k]<D_{\mathrm{th}}.
\end{cases}
\label{eq:event_activation}
\end{equation}

The common activation threshold is applied across network configurations using the normalized measurement representation. The sensitivity of this equal-weight formulation to individual measurement contributions is examined in the performance evaluation.

\subsection{Sequence-Admittance Indicators and Local Directional Classification}

While the disturbance index in Section~III-A quantifies local disturbance magnitude, directional discrimination requires additional sequence-domain information. Accordingly, two admittance indicators and a signed incremental directional index are employed to characterize the local terminal response.

The positive-sequence current-to-voltage indicator is defined as
\begin{equation}
\Psi_{\mathrm{sym}}[k]=
\frac{|I_1[k]|}
{\max\left(|V_1[k]|,V_{\min}\right)}.
\label{eq:sym_admittance}
\end{equation}
\noindent Here, $I_1[k]$ and $V_1[k]$ are the complex positive-sequence current and voltage components obtained from the sampled three-phase measurements, and $V_{\min}>0$ prevents numerical instability during voltage depressions.

The incremental positive-sequence components, referenced to their prefault values, are
\begin{equation}
\begin{aligned}
\Delta I_1[k]&=I_1[k]-I_{1,\mathrm{pre}},\\
\Delta V_1[k]&=V_1[k]-V_{1,\mathrm{pre}}.
\end{aligned}
\label{eq:incremental_components}
\end{equation}
The corresponding incremental admittance magnitude is
\begin{equation}
\Psi_{\Delta Y}[k]=
\begin{cases}
\dfrac{|\Delta I_1[k]|}
{|\Delta V_1[k]|},
& |\Delta V_1[k]|>V_{\Delta,\min},\\[5pt]
0, & \text{otherwise},
\end{cases}
\label{eq:incremental_admittance}
\end{equation}
\noindent where $V_{\Delta,\min}>0$ avoids division by near-zero incremental voltages. Its per-unit representation is obtained using
\begin{equation}
Y_{\mathrm{base}}=\frac{I_{\mathrm{base}}}{V_{\mathrm{base}}},
\qquad
\Psi_{\Delta Y,\mathrm{pu}}[k]
=\frac{\Psi_{\Delta Y}[k]}{Y_{\mathrm{base}}},
\label{eq:normalized_incremental_admittance}
\end{equation}
\noindent where $I_{\mathrm{base}}$ and $V_{\mathrm{base}}$ are the rated base quantities defined in Section~III-A.

To retain the relative phase information between incremental voltage and current, the signed directional index is defined as
\begin{equation}
\Gamma[k]=
\frac{
\operatorname{Re}\left\{
\Delta V_1[k]\Delta I_1^{*}[k]
\right\}
}{
|\Delta V_1[k]|
|\Delta I_1[k]|+\varepsilon
}.
\label{eq:directional_index}
\end{equation}
\noindent Here, $(\cdot)^*$ denotes complex conjugation, $\operatorname{Re}\{\cdot\}$ extracts the real component, and $\varepsilon>0$ is a numerical regularization constant with the same physical units as the denominator product.

The forward directional reference is determined by the relay current-polarity convention, with $s_{\mathrm{fwd}}\in\{-1,+1\}$ denoting its forward sign. The directional register is updated as
\begin{equation}
M[k]=
\begin{cases}
0, & s_{\mathrm{fwd}}\Gamma[k]>\gamma_{\mathrm{dir}}
\ \land\ D[k]>D_{\mathrm{th}},\\
1, & s_{\mathrm{fwd}}\Gamma[k]<-\gamma_{\mathrm{dir}}
\ \land\ D[k]>D_{\mathrm{th}},\\
M[k-1], & \text{otherwise},
\end{cases}
\label{eq:directional_register}
\end{equation}
\noindent where $M[k]=0$ and $M[k]=1$ denote forward and reverse states, respectively, and $\gamma_{\mathrm{dir}}\geq0$ is the directional security margin. Within the indeterminate region, the previous state is retained; the initial state and reset conditions are specified during relay implementation.

The directional register supervises neuronal firing, permitting tripping only when the disturbance criterion and forward-direction condition are satisfied. The directional polarity and security margin are established according to relay orientation and evaluated under internal and external faults, load variations, and converter-limited operating conditions.

\subsection{Event-Driven Spike Encoding and Direction-Dependent Neuronal Firing}

The proposed framework employs event-driven spike encoding to transform the disturbance index into neuronal excitation while suppressing unnecessary spike generation during steady-state operation. The disturbance activation signal is defined as
\begin{equation}
S_{\mathrm{act}}[k]=
\begin{cases}
1, & D[k]\geq D_{\mathrm{th}},\\
0, & \text{otherwise}.
\end{cases}
\label{eq:disturbance_activation}
\end{equation}

A new disturbance event is detected from the rising edge of the activation signal:
\begin{equation}
E_{\mathrm{on}}[k]
=
S_{\mathrm{act}}[k]\land
\neg S_{\mathrm{act}}[k-1].
\label{eq:event_onset}
\end{equation}
At each event onset $k_e$, the instantaneous disturbance magnitude is sampled to determine the encoding latency:
\begin{equation}
\tau_{\mathrm{lat},e}
=
\max\left(
\frac{\tau_{\max}}
{1+\kappa D[k_e]/D_{\mathrm{th}}},
2\Delta t_{\mathrm{sim}}
\right),
\label{eq:spike_latency}
\end{equation}
where $\tau_{\max}$ denotes the maximum encoding latency, $\Delta t_{\mathrm{sim}}$ is the simulation sampling interval, and $\kappa>0$ controls the sensitivity of latency to disturbance magnitude. The calculated latency is held constant throughout the corresponding event, while its lower bound ensures a minimum duration of two simulation steps.

For a prescribed maximum disturbance index $D_{\max}$, the sensitivity coefficient is calibrated as
\begin{equation}
\kappa=
\frac{D_{\mathrm{th}}}{D_{\max}}
\left(
\frac{\tau_{\max}}{2\Delta t_{\mathrm{sim}}}-1
\right).
\label{eq:latency_sensitivity}
\end{equation}
This calibration ensures that the encoding latency reaches its minimum value of $2\Delta t_{\mathrm{sim}}$ when $D[k_e]=D_{\max}$, provided that $D_{\max}>D_{\mathrm{th}}$ and $\tau_{\max}>2\Delta t_{\mathrm{sim}}$. Consequently, larger disturbance magnitudes result in shorter encoding latencies, enabling a faster neuronal response to more severe disturbances.

A presynaptic spike is generated when the elapsed time from the event onset reaches the calculated latency:
\begin{equation}
S_{\mathrm{spike}}[k]=
\begin{cases}
1, & S_{\mathrm{act}}[k]=1,\ 
(k-k_e)\Delta t_{\mathrm{sim}}\geq\tau_{\mathrm{lat},e},\\
0, & \text{otherwise}.
\end{cases}
\label{eq:presynaptic_spike}
\end{equation}
Here, $k_e$ denotes the event-onset sample, and $S_{\mathrm{spike}}[k]$ represents a single-sample presynaptic spike. A generation flag prevents repeated spikes during the same activation event and is reset when the event terminates. Thus, each sustained disturbance event produces at most one presynaptic spike. The resulting neuronal input is expressed as
\begin{equation}
I_{\mathrm{spike}}[k]
=
I_0 S_{\mathrm{spike}}[k],
\label{eq:spike_input}
\end{equation}
where $I_0$ denotes the prescribed spike-current amplitude.

The neuronal input drives the membrane dynamics according to the discrete-time leaky integrate-and-fire (LIF) model:
\begin{equation}
V_m^{*}[k]=
V_m[k-1]e^{-\Delta t_{\mathrm{sim}}/\tau_{\mathrm{leak}}}
+
\frac{R_{\mathrm{syn}}\Delta t_{\mathrm{sim}}}
{\tau_{\mathrm{mem}}}I_{\mathrm{spike}}[k].
\label{eq:lif_discrete_dynamics}
\end{equation}
Here, $V_m[k-1]$ is the membrane potential retained from the preceding sampling instant, whereas $V_m^{*}[k]$ denotes the updated potential before reset. The parameters $\tau_{\mathrm{leak}}$ and $\tau_{\mathrm{mem}}$ represent the leakage and membrane integration time constants, respectively, while $R_{\mathrm{syn}}$ denotes the synaptic resistance. These time constants are related through
\begin{equation}
\tau_{\mathrm{mem}}=RC,
\qquad
\tau_{\mathrm{leak}}=K_{\mathrm{leak}}\tau_{\mathrm{mem}},
\label{eq:membrane_parameters}
\end{equation}
where $R$ and $C$ are the effective membrane resistance and capacitance, respectively, and $K_{\mathrm{leak}}$ determines the leakage timescale. This formulation enables the membrane to accumulate successive neuronal excitations while allowing the stored potential to decay exponentially between excitation events.

To incorporate directional selectivity into neuronal firing, the adaptive threshold is determined using the preceding membrane state and the instantaneous directional classification:
\begin{equation}
\Theta[k]=
\begin{cases}
V_{\mathrm{rest}}+V_m[k-1], & M[k]=1,\\[4pt]
V_{\mathrm{rest}}e^{-\lambda_{\mathrm{fast}}D[k]/D_{\mathrm{th}}}
+\dfrac{1}{2}V_m[k-1], & M[k]=0.
\end{cases}
\label{eq:adaptive_firing_threshold}
\end{equation}
Here, $V_{\mathrm{rest}}$ denotes the resting membrane potential, and $\lambda_{\mathrm{fast}}>0$ controls the disturbance-dependent threshold reduction. The reverse-direction condition imposes a higher firing threshold to restrain unwanted operation, whereas the forward-direction condition progressively reduces the threshold as the disturbance magnitude increases, facilitating faster neuronal firing for stronger disturbances.

The neuronal firing output is obtained by comparing the updated membrane potential with the corresponding adaptive threshold:
\begin{equation}
F[k]=
\begin{cases}
1, & V_m^{*}[k]\geq\Theta[k],\\
0, & \text{otherwise}.
\end{cases}
\label{eq:neuronal_firing}
\end{equation}

A trip command is generated only when neuronal firing coincides with forward-direction permission and an active disturbance condition. Accordingly, the final protection logic is given by
\begin{equation}
\mathrm{Trip}[k]
=
F[k]\land\big(M[k]=0\big)\land S_{\mathrm{act}}[k].
\label{eq:trip_logic}
\end{equation}

Following the protection decision, the membrane potential is updated through a subtractive reset:
\begin{equation}
V_m[k]=
V_m^{*}[k]-\mathrm{Trip}[k]\Theta[k].
\label{eq:partial_reset}
\end{equation}
This operation subtracts the firing threshold when a trip command is issued while retaining the residual membrane excitation for subsequent neuronal integration.

The resulting protection response is governed by the combined effects of disturbance-dependent encoding latency, temporal membrane integration, adaptive threshold modulation, and directional classification. Their influence on fault detection and protection performance is examined under different network and operating conditions in Section~IV.

\section{Real-Time Validation and Protection Performance}

\begin{figure}[t]
\centering
\hspace*{-0.8cm}
\includegraphics[scale=1.1]{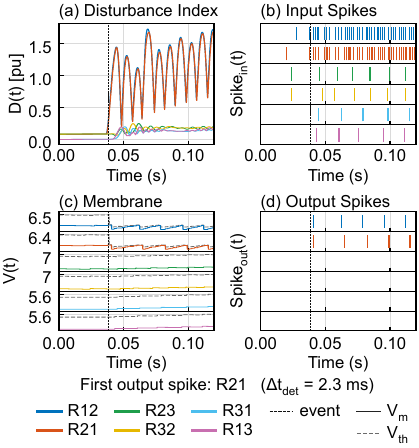}
\caption{Temporal signatures for an AG fault (islanded ring) – in-zone relays R12, R21 trip; others restrained.}
\label{fig3}
\end{figure}

\begin{figure}[t]
\centering
\hspace*{-0.8cm}
\includegraphics[scale=1.1]{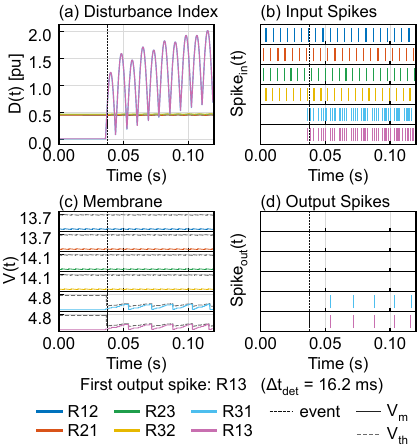}
\caption{Temporal signatures for a BC fault (islanded ring) – in-zone relays R31, R13 trip; others restrained.}
\label{fig4}
\end{figure}

\begin{figure}[t]
\centering
\hspace*{-0.8cm}
\includegraphics[scale=1.1]{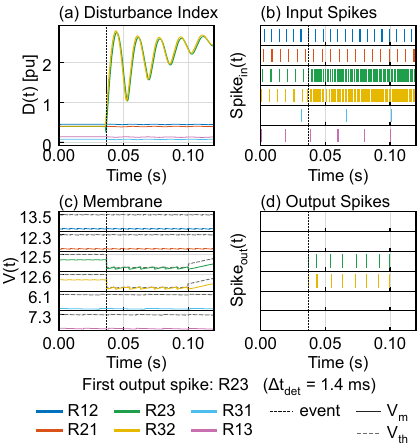}
\caption{Temporal signatures for an ABCG fault (islanded ring) – in-zone relays trip within one cycle; out-of-zone relays restrained.}
\label{fig5}
\end{figure}

\begin{table}[!t]
    \renewcommand{\arraystretch}{1.12}
    \caption{Validation Network Configurations}
    \label{tab1}
    \centering
    \footnotesize
    \begin{tabular}{@{}lcc@{}}
        \toprule
        \textbf{Parameter} & \textbf{Islanded ring} & \textbf{IEEE 13-node}\\
        \midrule
        Nominal voltage (L--L) & 480 V & 4.16 kV\\
        Frequency & 60 Hz & 60 Hz\\
        Base power / rating & 100 kVA/DER & Benchmark feeder\\
        Topology & Closed-loop ring & Mixed ring--radial\\
        Operating mode & Islanded & Grid-connected\\
        \midrule
        \multicolumn{3}{@{}l@{}}{\textit{Shared constants:}
        $k=400,\quad U_r=1.0,\quad \lambda=0.5$}\\
        \bottomrule
    \end{tabular}
\end{table}

\subsection{Real-Time Test Systems and Implementation}

The proposed protection framework is evaluated on two distribution networks: an IEEE 13-node radial feeder and a three-DER islanded ring microgrid, as shown in Fig.~\ref{fig2}. Both systems are implemented on the RTDS-NOVA COR platform using RSCAD FX 2.6, with a simulation time step of $50~\mu\mathrm{s}$. Protection decisions rely exclusively on local electrical measurements, without exchanging remote voltage or current signals. The network ratings, configurations, and operating conditions are summarized in Table~\ref{tab1}.

The islanded ring microgrid comprises three interconnected line sections protected by six relays arranged in three opposing pairs: $(R_{12},R_{21})$, $(R_{23},R_{32})$, and $(R_{31},R_{13})$. Its multiple fault-current paths provide a suitable environment for assessing directional discrimination. The IEEE 13-node feeder operates at 4.16~kV and includes voltage regulation and shunt compensation, with three relays monitoring selected feeder sections: RG60 (650--632), BUS632 (632--671), and BUS633 (632--633).

The evaluation considers AG, BC, and ABCG faults at different locations in the islanded ring, and AG, ABG, and A-D-phase faults at selected locations in the IEEE 13-node feeder. Sudden load changes are also introduced in both networks to assess protection security under non-fault conditions. The performance assessment examines fault dependability, directional selectivity, restraint of non-target relays, and operating time, measured from fault inception to relay trip assertion.

\subsection{Fault Scenarios and Directional Selectivity}

\begin{figure}[t]
\centering
\hspace*{-0.8cm}
\includegraphics[scale=1.1]{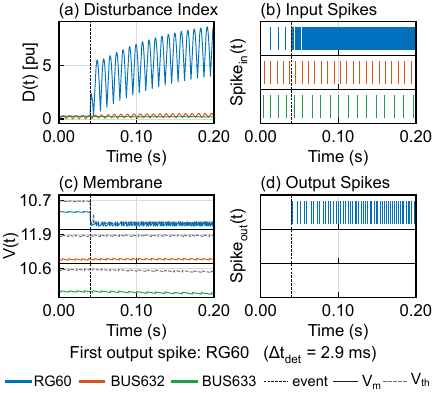}
\caption{Temporal signatures for an AG fault at node 650 (IEEE 13-node) – RG60 operates first; downstream relays restrained.}
\label{fig6}
\end{figure}

\begin{figure}[t]
\centering
\hspace*{-0.8cm}
\includegraphics[scale=1.1]{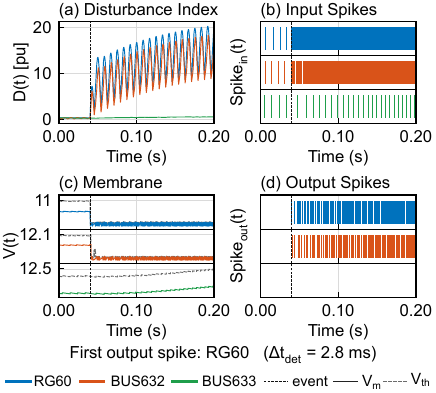}
\caption{Temporal signatures for an AB fault at node 633 (IEEE 13-node) – in-zone relays operate; others restrained.}
\label{fig7}
\end{figure}

\begin{figure}[t]
\centering
\hspace*{-0.8cm}
\includegraphics[scale=1.1]{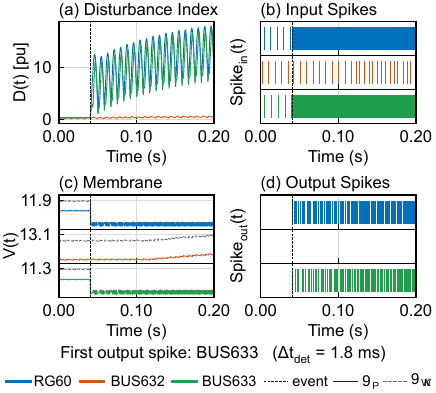}
\caption{Temporal signatures for an ABG fault at node 632 (IEEE 13-node) – RG60 and BUS632 operate; BUS633 restrained.}
\label{fig8}
\end{figure}

\begin{figure}[t]
\centering
\hspace*{-0.8cm}
\includegraphics[scale=1.1]{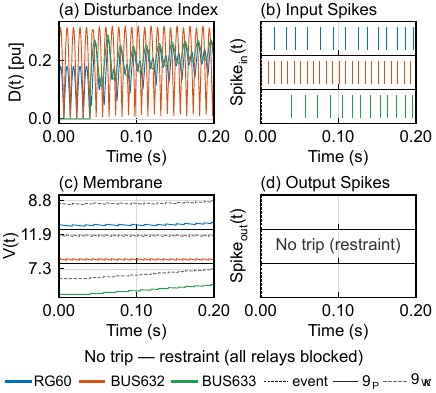}
\caption{Protection security under a load change (IEEE 13-node) – no relay trips, confirming restraint against non-fault disturbances.}
\label{fig9}
\end{figure}

Directional selectivity is evaluated through multiple fault scenarios across the protected line sections of the islanded ring microgrid and the IEEE 13-node distribution feeder. The corresponding scenarios assess fault detection, first output-spike response, and restraint of out-of-zone relays.

\textbf{1) Islanded Ring Microgrid:}

\textbf{Case I---AG Fault on Line 1--2:} An A-phase-to-ground (AG) fault is initiated at $t=0.0400~\mathrm{s}$. As shown in Fig.~\ref{fig3}, relays $R_{12}$ and $R_{21}$ respond, while the remaining relays remain restrained. The first output spike occurs at $R_{21}$ after 2.3~ms, demonstrating rapid detection and selective operation despite multiple fault-current paths in the ring.

\textbf{Case II---BC Fault on Line 1--3:} A B--C phase-to-phase (BC) fault is applied at $t=0.0400~\mathrm{s}$. Fig.~\ref{fig4} shows that relays $R_{31}$ and $R_{13}$ respond, with the out-of-zone relays remaining restrained. The first output spike occurs at $R_{13}$ after 16.2~ms. The longer detection time compared with the AG fault reflects differences in the electrical signatures and their temporal evolution through spike encoding and neuronal integration.

\textbf{Case III---ABCG Fault on Line 2--3:} A three-phase-to-ground (ABCG) fault is initiated at $t=0.0400~\mathrm{s}$. As illustrated in Fig.~\ref{fig5}, relays $R_{23}$ and $R_{32}$ respond, while the out-of-zone relays remain restrained. The first output spike occurs at $R_{23}$ after 1.4~ms, representing the fastest response among the evaluated scenarios due to the pronounced electrical disturbance.

Across the three fault scenarios, first output-spike detection times range from 1.4 to 16.2~ms. In each case, the corresponding in-zone relay pair responds while the remaining relays are restrained, demonstrating directional selectivity across all three protected line sections.

\textbf{2) IEEE 13-Node Distribution Feeder:}

The IEEE 13-node distribution feeder is subjected to three fault scenarios and one non-fault disturbance to assess the directional selectivity of the proposed protection framework across different feeder locations and protection zones.

\textbf{Case I---AG Fault at Node 650:} An AG fault is initiated at $t=0.0400~\mathrm{s}$. As shown in Fig.~\ref{fig6}, RG60 responds first, while the downstream relays remain restrained. The first output spike is detected at RG60 after $2.9~\mathrm{ms}$, indicating a rapid response to the fault at the monitored feeder location without unintended operation of the other relays.

\textbf{Case II---AB Fault at Node 633:} An AB fault is applied at Node 633 at $t=0.0400~\mathrm{s}$. As illustrated in Fig.~\ref{fig7}, the in-zone relays respond, whereas the remaining relays remain restrained. The first output spike is observed at RG60 after $2.8~\mathrm{ms}$, demonstrating the response of the directional protection mechanism to a phase-to-phase fault at a different feeder location.

\textbf{Case III---ABG Fault at Node 632:} An ABG fault is initiated at $t=0.0400~\mathrm{s}$. As shown in Fig.~\ref{fig8}, RG60 and BUS632 operate, while BUS633 remains restrained. The first output spike is reported at BUS633 after $1.8~\mathrm{ms}$. This scenario evaluates the framework's response to a double-line-to-ground fault and its ability to discriminate between the monitored feeder sections.

\begin{figure}[t]
  \centering
  \includegraphics[width=\linewidth]{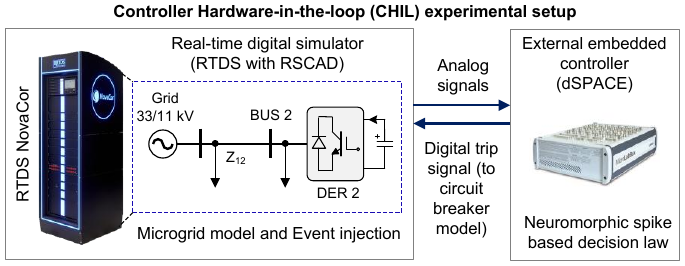}
  \caption{Controller-hardware-in-the-loop (CHIL) test setup (RTDS–dSPACE) with the neuromorphic protection controller.}
 \label{fig10}
\end{figure}

\textbf{Case IV---Sudden Load Addition at Bus 634:} A load of $160~\mathrm{kW}+j110~\mathrm{kVAr}$ is connected at Bus~634 at $t=0.0400~\mathrm{s}$. Despite the resulting transient variations in the measured electrical quantities, none of the monitored relays issues a trip command, as shown in Fig.~\ref{fig9}. This response demonstrates the framework's ability to maintain protection security during the tested load-change condition. Unlike a fault, the load transient does not satisfy the combined disturbance-activation, directional-supervision, and neuronal-firing conditions required to initiate a trip.

\section{Hardware-in-the-Loop Validation and Performance Characterization}

Whereas Section~IV established in-zone sensitivity and out-of-zone security through individual fault and non-fault scenarios, this section addresses two complementary questions that are not tied to a single scenario: whether the framework executes correctly on physical protection hardware, and how its operating time varies statistically across fault types, network configurations, and fault-resistance levels. The former is examined through controller-hardware-in-the-loop (CHIL) experiments, and the latter through a quantitative performance study over $N=240$ cases.

\subsection{Controller-Hardware-in-the-Loop Validation}

The proposed protection framework was evaluated using a CHIL platform comprising an RTDS real-time simulator and an external dSPACE controller. The three-phase voltages $(V_a,V_b,V_c)$ and breaker currents $(I_a,I_b,I_c)$ generated by the RTDS were fed to the analog inputs of the dSPACE controller. The neuromorphic protection algorithm was executed on the dSPACE processor, and the resulting binary trip command was transmitted through a digital output to an RTDS digital input. This interface enabled real-time signal acquisition, hardware-based protection processing, and trip-command feedback, as illustrated in Fig.~\ref{fig10}. The circuit breaker was kept disabled throughout the experiments; accordingly, the reported operating times correspond to trip-signal assertion and do not represent physical breaker operation or fault-current interruption.

\begin{figure}[h]
  \centering
  \includegraphics[width=0.9\columnwidth]{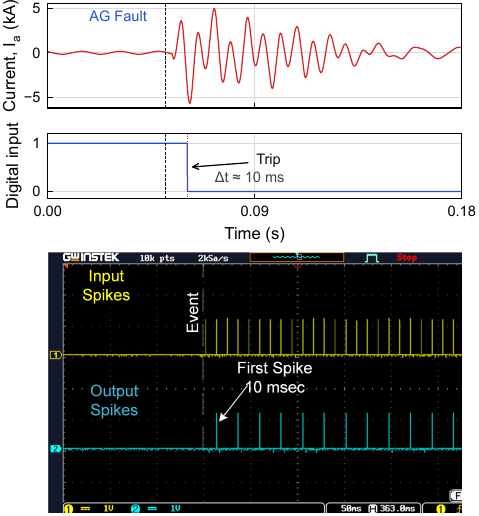}
  \caption{CHIL validation for an AG fault – grid current, hardware trip signal ($\Delta t\approx10$~ms), and input/output spike capture.}
  \label{fig11}
\end{figure}

\begin{figure}[h]
  \centering
  \includegraphics[width=0.9\columnwidth]{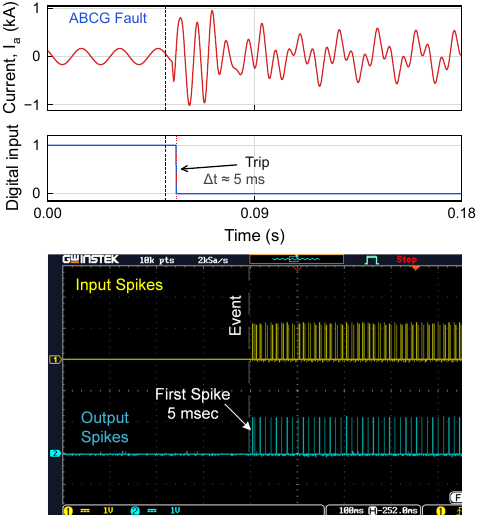}
  \caption{CHIL validation for an ABCG fault – grid current, hardware trip signal ($\Delta t\approx5$~ms), and spike capture.}
  \label{fig12}
\end{figure}

\begin{figure}[h]
  \centering
  \includegraphics[width=0.9\columnwidth]{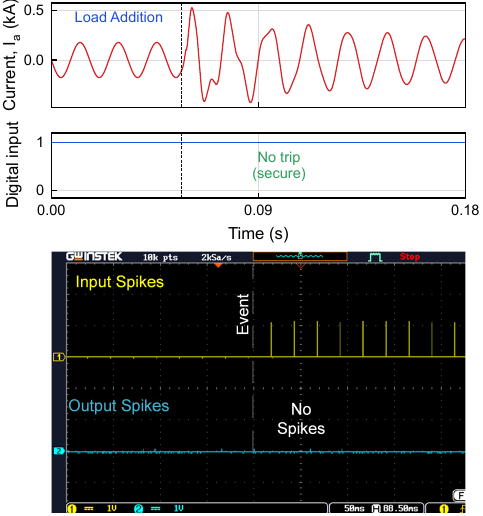}
  \caption{CHIL security under a load addition – no trip; output shows “No spikes”.}
  \label{fig13}
\end{figure}

\begin{table}[!t]
\centering
\caption{Comparison with recent microgrid protection methods}
\label{tab2}
\renewcommand{\arraystretch}{1.2}
\setlength{\tabcolsep}{3pt}
\scriptsize
\begin{tabular}{@{}p{2.7cm}p{2.0cm}p{3.0cm}@{}}
\toprule
\textbf{Approach} & \textbf{Operating time} & \textbf{Validation / notes} \\
\midrule
\textbf{Proposed neuromorphic (this work)} & \textbf{1--4\,ms grid; 8--9\,ms islanded} & \textbf{Real-time CHIL; passive, 95\% $\le$20\,ms} \\
\addlinespace[3pt]
ROCOF $d$-axis~\cite{farshadi2023rate} & Sub-cycle & Simulation; local, passive \\
\addlinespace[3pt]
Decision-tree ML~\cite{beikbabaei2024machine} & $<5$\,ms & Simulation; needs training set \\
\addlinespace[3pt]
Harmonic injection~\cite{khan2022communication} & Time-graded & HIL; tied to inverter control \\
\addlinespace[3pt]
Neg.-seq.\ injection~\cite{yang2024negative} & $\approx$140\,ms + delay & Simulation; slow injection \\
\bottomrule
\end{tabular}
\end{table}

\begin{figure}[h]
  \centering
  \includegraphics[width=0.95\columnwidth]{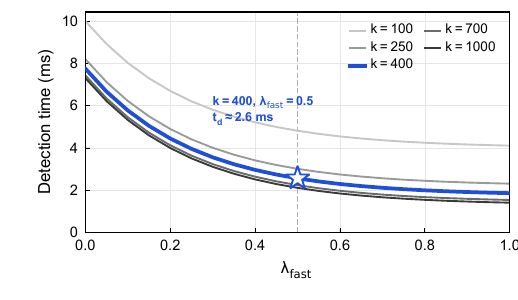}
  \caption{Sensitivity of detection time to \(\lambda_{\mathrm{fast}}\) for different values of \(k\). The selected operating point (\(\lambda_{\mathrm{fast}}=0.5\), \(t_d\approx2.6\) ms) is marked by a star.}
  \label{fig14}
\end{figure}

\textbf{1) Response to Internal Faults:} Two internal-fault scenarios were considered. First, an A-phase-to-ground (AG) fault was applied with the DER operating at its rated capacity of 100~kVA. As shown in Fig.~\ref{fig11}, the current transient and voltage reduction were followed by hardware trip-signal assertion after approximately $10~\mathrm{ms}$, and the input- and output-spike records confirm execution of the neuromorphic processing sequence on the external controller. A solid three-phase-to-ground (ABCG) fault was then introduced with the DER rating reduced to 50~kVA to assess the response under a weaker-source condition. As illustrated in Fig.~\ref{fig12}, the hardware trip signal was asserted within approximately $5~\mathrm{ms}$, accompanied by increased input-spike activity and rapid output-spike generation.

\textbf{2) Response to a Non-Fault Disturbance:} Protection security was evaluated by connecting an inductive load corresponding to 40\% of the rated microgrid capacity while the DER operated at 100~kVA. Although the load connection caused a transient increase in phase current, Fig.~\ref{fig13} shows that no output spikes or trip command were generated, and the membrane potential remained below the adaptive firing threshold.

\begin{figure*}[h]      
  \centering
  \includegraphics[width=0.95\textwidth]{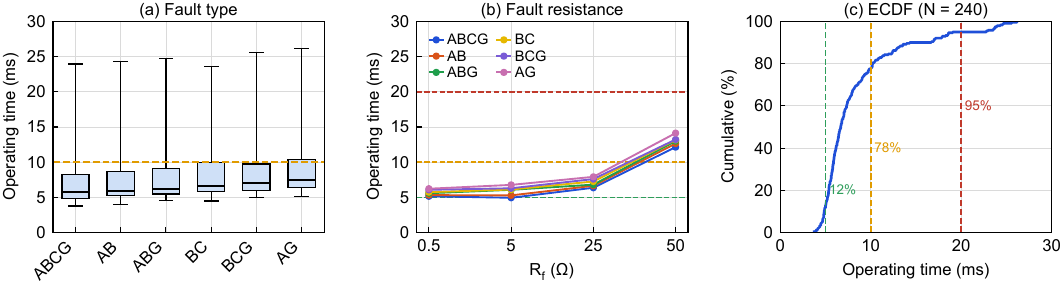}
  \caption{Quantitative protection performance over the two networks ($N=240$):
  (a) operating-time distribution by fault type, (b) operating time versus fault
  resistance, and (c) cumulative operating-time distribution.}
  \label{fig15}
\end{figure*}

The CHIL experiments verify hardware execution of the proposed protection algorithm, with trip-signal assertion at approximately 10~ms and 5~ms for the AG and ABCG faults, respectively, and correct restraint under the tested non-fault disturbance. Since the circuit breaker remained disabled, the validation is limited to protection decision-making and trip-command generation rather than physical fault clearance. The proposed framework is further compared with recent microgrid protection methods in Table~\ref{tab2}.

\subsection{Parameter Sensitivity and Quantitative Protection Performance}

The influence of the adaptive parameters $\lambda_{\mathrm{fast}}$ and $k$ on detection time is first examined, as shown in Fig.~\ref{fig14}. Detection time decreases with increasing $\lambda_{\mathrm{fast}}$, with the reduction rate depending on the selected $k$. The selected operating point, indicated by a star, is $\lambda_{\mathrm{fast}}=0.5$ and $k=400$, yielding a detection time of approximately $2.6~\mathrm{ms}$. These values are retained for the subsequent performance evaluation.

The operating-time performance is evaluated across both networks using $N=240$ test cases (Fig.~\ref{fig15}). The median operating times remain below the 10-ms reference across all fault categories, although some upper-tail observations exceed this limit (Fig.~\ref{fig15}(a)). The medians range from 5.4 to 7.1~ms in the islanded ring and from 6.2 to 7.6~ms in the IEEE~13-node feeder, indicating a slightly slower response in the latter.

Operating time increases with fault resistance across all six fault categories, rising from approximately 5--6~ms at $R_f=0.5~\Omega$ to 12--14~ms at $R_f=50~\Omega$ (Fig.~\ref{fig15}(b)). This trend reflects the weaker fault-induced disturbances at higher resistance, which require longer neuronal integration to reach the firing threshold. The cumulative distribution further indicates that approximately 12\%, 78\%, and 95\% of the evaluated cases are resolved within 5, 10, and 20~ms, respectively (Fig.~\ref{fig15}(c)). These results characterize the protection response across different fault types, network configurations, and fault-resistance levels.

\section{Conclusion}

This paper proposed a directionally selective neuromorphic protection framework that integrates disturbance-responsive leaky integrate-and-fire dynamics with local sequence-domain directional discrimination to achieve communication-free protection. By combining temporal variations in local measurements with directional information, the framework enables selective fault tripping without remote electrical measurements or inter-relay communication.

Real-time RTDS validation on an islanded ring microgrid and an IEEE 13-node feeder demonstrated selective responses across the evaluated fault scenarios, with no unintended tripping during the tested load addition. Across 240 cases, median operating times ranged from 5.4--7.1~ms in the islanded ring and 6.2--7.6~ms in the IEEE 13-node feeder; 78\% of cases were resolved within 10~ms. Operating times increased from approximately 5--6~ms to 12--14~ms as fault resistance increased from 0.5 to 50~$\Omega$. CHIL experiments further demonstrated hardware-based trip-signal assertion within approximately 10~ms for an AG fault and 5~ms for an ABCG fault, with no trip during the tested load addition. Collectively, these results demonstrate the effectiveness of the proposed communication-free neuromorphic framework in providing fast and selective protection across heterogeneous microgrid configurations, with successful real-time and hardware-in-the-loop validation.





\footnotesize{\bibliographystyle{IEEEtran}}

\bibliography{ref}


\end{document}